\documentclass[9pt,twocolumn,twoside]{pnas-new}
\usepackage{hyperref}
\usepackage{bm}
\templatetype{pnasresearcharticle} 

\setboolean{displaywatermark}{false}
\leadauthor{Mori} 

\usepackage{fancyhdr}
\fancypagestyle{firststyle}{
 \fancyfoot[R]{\textbf{\thepage}}
 \fancyfoot[L]{}
 \fancyhead[R]{}
 \fancyhead[L]{}
}

\title{Cities and Space: Common Power Laws and Spatial Fractal Structures}

\author[a,b,1]{Tomoya Mori}
\author[c]{Tony E. Smith} 
\author[d]{Wen-Tai Hsu}

\affil[a]{Institute of Economic Research,
Kyoto University. Yoshida-Honmachi, Sakyo-Ku, Kyoto, 606-8501 Japan.
Phone: +81-75-753-7121, E-mail: mori@kier.kyoto-u.ac.jp.}
\affil[b]{Research Institute of Economy, Trade and Industry (RIETI), 11th floor, Annex, Ministry of Economy, Trade and Industry (METI) 1-3-1, Kasumigaseki Chiyoda-ku, Tokyo 100-8901, Japan.}
\affil[c]{Department of Electrical and Systems Engineering, University
of Pennsylvania, PA 19104, USA. Phone: +1-215-898-9647, E-mail: tesmith@seas.upenn.edu.}
\affil[d]{School of Economics, Singapore Management University.
90 Stamford Road, Singapore 178903. Phone: +65-6808-5455, E-mail:
wentaihsu@smu.edu.sg.}

\significancestatement{%
Socio-economic attributes of cities (e.g., wages, education, GDP, industrial diversity, crime) have been shown to exhibit strong correlations with city size as measured by population.
It has thus been a major research objective to characterize and explain city size distributions.
Whereas city size distributions are known to exhibit power laws at the country level, we find that they exhibit strikingly similar power laws when examined along a spatial hierarchy of regions.
Such high degree of similarity could not be obtained if city sizes were generated by a random (growth) process.
The fact that this regularity emerges along spatial dimensions in a recursive manner suggests the existence of spatial fractal structures and calls for new theories to explain their underlying mechanisms.}

\authorcontributions{Author contributions: T.M. developed the ideas of the paper and conducted analyses. 
T.E.S. formalized the concepts and developed estimation framework and test statistics.
W.H. helped refine the concepts and took the leading role in organizing the manuscript.
All authors contributed designing the hypothesis tests, interpreted the test results and the preparation of the manuscript.}
\authordeclaration{The authors declare no conflict of interest.}
\correspondingauthor{\textsuperscript{1}To whom correspondence should be addressed. E-mail: mori@\@kier.kyoto-u.ac.jp}


\begin{abstract}
City size distributions are known to be well approximated by power laws across a wide range of countries. 
But such distributions are also meaningful at other spatial scales, such as within certain regions of a country. Using data from China, France, Germany, India, Japan, and the US, we first document that large cities are significantly more spaced out than would be expected by chance alone. 
We next construct spatial hierarchies for countries by first partitioning geographic space using a given number of their largest cities as cell centers, and then continuing this partitioning procedure within each cell recursively. 
We find that city size distributions in different parts of these spatial hierarchies exhibit power laws that are again far more similar than would be expected by chance alone -- suggesting the existence of a spatial fractal structure.
\end{abstract}

\dates{This manuscript was compiled on \today}
\doi{\url{www.pnas.org/cgi/doi/10.1073/pnas.XXXXXXXXXX}}

\begin{document}

\maketitle
\thispagestyle{firststyle}
\ifthenelse{\boolean{shortarticle}}{\ifthenelse{\boolean{singlecolumn}}{\abscontentformatted}{\abscontent}}{}

\dropcap{A} variety of power-law properties related to cities (both within and
across cities) have been documented \cite{Batty-2006,Bettencourt-et-al-2007,Bettencourt-2013,Bettencourt-West-2010}.
In particular, city size distributions are known to be well approximated
by power laws across a wide range of countries \cite{Gabaix-Ioannides-2004}.
But one may also examine city size distributions at other spatial scales,
such as within certain regions of a country. 
A natural question is whether there is any relation among city-size distributions in different spatial units.
One possibility is related to the idea of fractal structure, in which smaller parts of a system structurally resemble the larger ones, including the entire system \cite{Mandelbrot-1982}. 
If any system is a fractal structure and exhibits a power law as a whole, then the scale-invariant property of fractal structures implies that its smaller parts must also exhibit similar power laws. More generally, whenever a system exhibits this similarity property, the system is said to exhibit a \emph{common power law} (CPL).

Examples of fractal structures are diverse, from biology \cite{West-et-al-1997,Ravasz-et-al-2002} to the internet \cite{Faloutsos-et-al-1999,Yook-et-al-2002} to firms
\cite{Stanley-et-al-1996} and cities \cite{Batty-Longley-1994,Batty-2013,Makse-et-al-1995}.
With respect to cities in particular, there is some empirical evidence to suggest that individual cities can be viewed as fractal structures \cite{Batty-Longley-1994,Makse-et-al-1995,Rybski-Ros-PHYR-E2013}.
But is this also true of the entire system of cities within a country?
This article provides the first evidence of striking similarities among city size distributions in terms of their power laws when such city systems are viewed as spatial hierarchies.
This spatially oriented CPL result suggests the existence of spatial fractal structure at the city-system level. 
\begin{figure*}[t]
\centering{}\includegraphics[width=17.8cm,height=11.657cm]{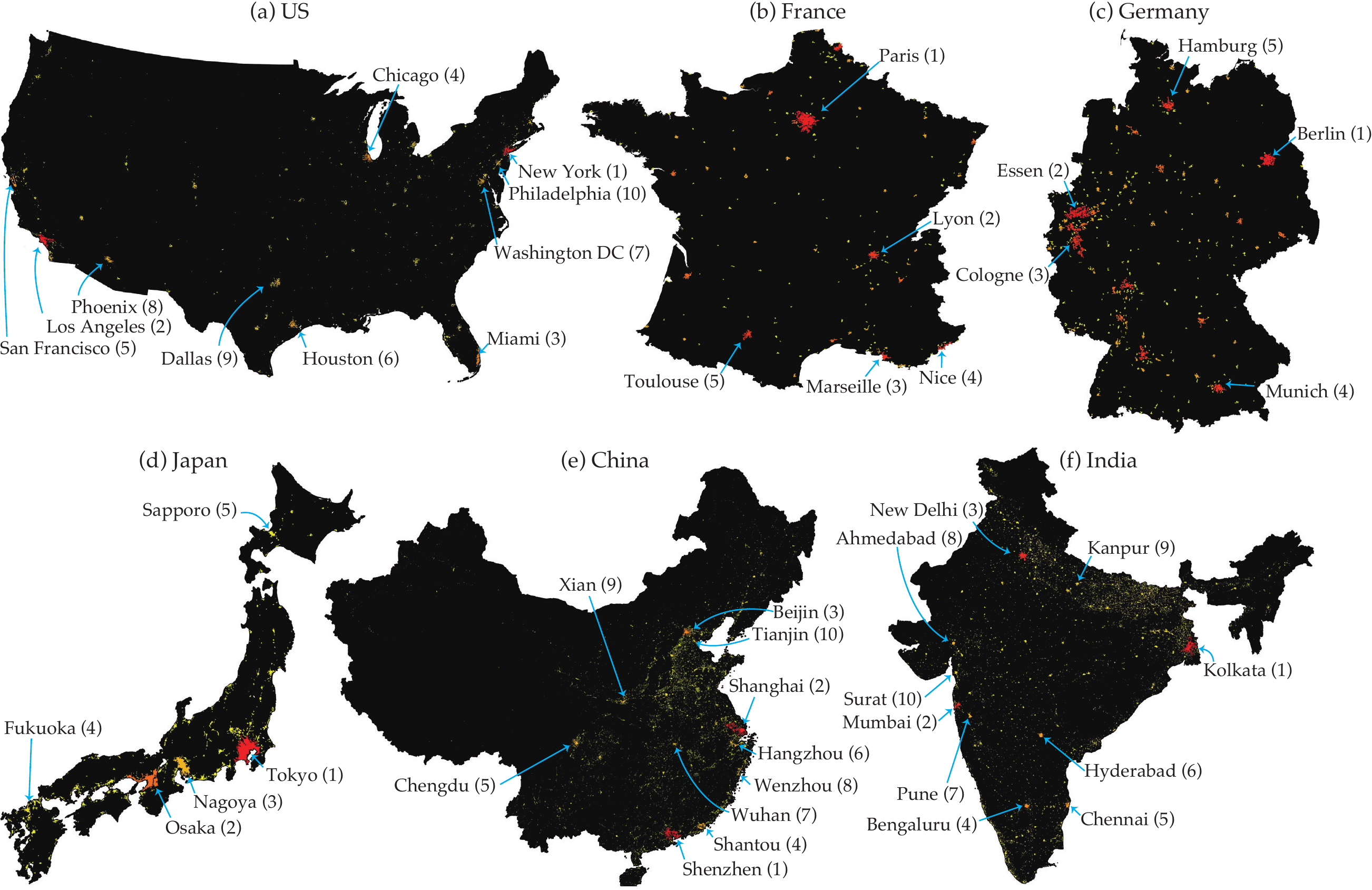}
\vspace{5pt}
 \caption{Cities in the US, France, Germany, Japan, China and India are shown on the corresponding maps, where the numbers of cities are 931, 202, 331, 450, 7,203 and 7,913, respectively. In each panel, non-black areas indicate cities, where a warmer color corresponds to a larger population size. For each of the three larger countries, the US, China, and India, the largest ten cities are indicated with their population rankings, whereas the largest five cities are indicated for the other three countries.\label{fig:ua}}
\end{figure*}

The most popular theoretical derivation of power laws for city-size distributions postulates that growth rates of individual cities are independently and identically distributed (\emph{iid}) random variables \cite{Gabaix-1999,Duranton-2006,Rossi-Hansberg-Wright-2007}, i.e., Gibrat's law \cite{Gibrat-1931}. 
This fundamental assumption necessarily implies that growth rates for any subset of these cities must also be \emph{iid}, and thus that the city system must have a fractal structure in the above sense. Moreover, the argument leading to a power law for the entire system must imply the same power law for each (sufficiently large) subset of cities, and thus must imply that this system exhibits a CPL. But this result is so inclusive that a CPL must hold for arbitrary subsets of cities, regardless of the spatial relations between them. In short, these random growth models suggest that spatial relations among cities do \emph{not} influence the distribution of city sizes.

However, there is a growing literature showing that space does indeed play a crucial role in shaping the economic landscape we observe.
At the city-system scale, distances between cities have been shown to influence both commodity flows and interactions between cities \cite{Duranton-et-al-2013,Redding-Sturm-2008}. 
At the within-city scale, distances between city centers and suburbs have been shown to influence a variety of urban phenomena (e.g., land use, housing, commuting patterns, and city growth) \cite{Anas-et-al-1998,Ahlfeldt-et-al-2015,Rosenfeld-et-al-2008}

Taken together, these many research efforts suggest that the distribution of city sizes may indeed be influenced by the spatial relations among these cities. 
To study this question, we begin by postulating that the spatial organization and sizes of cities are linked by the \emph{spatial grouping property} that larger cities tend to serve as centers around which smaller cities are grouped.
Moreover, this relation is recursive in the sense that some of these smaller cities may also serve as centers around which even smaller cities are grouped. 
For city landscapes that exhibit this type of hierarchical spatial grouping property, one might then expect to find similar city-size relations among groups. 
This in turn suggests that the CPL property above may indeed be stronger for such groupings than for arbitrary subsets of cities.

Given this line of reasoning, our main objective is to develop explicit tests of these hypotheses using data of city size and road distances for various countries. 
We first test one implication of the spatial grouping property which we call the \emph{spacing-out property}: the largest cities are spaced out relative to the whole set of cities.
We then construct appropriate hierarchical systems of sets and subsets of cities that are consistent with the spatial grouping property. A city system exhibiting both spatial grouping and a CPL will be said to exhibit a \emph{spatial CPL}.
By generating random counterfactual systems that differ only in terms of this spatial grouping property, we are able to conduct a \emph{spatial CPL test}: whether the power laws are significantly more similar in the systems that reflect spatial grouping relative to the random counterparts.

We find strong evidence for both the spacing-out property and spatial CPL property in essentially all countries tested.
Recall moreover that \emph{iid} random growth processes can also generate similar power laws across arbitrary (large) subsets of cities \cite{Gabaix-1999}, and that our random counterfactual systems are precisely collections of such subsets.
Thus the much tighter CPL result under systems that reflect spatial grouping implies that such a high degree of similarity could not be obtained if city sizes were generated by a random growth process.
We discuss various theoretical possibilities for explaining this spatial CPL property in the conclusion.

\section*{Data}

We examine countries that are relatively large in terms of both population and land area, and two groups of countries are considered. 
For countries in which the process of urbanization has essentially been completed, we consider the US, France, Germany, and Japan; for countries in which urbanization is still ongoing, we consider China and India.
We view cities as agglomerations of population, and employ the same definition of cities across countries. In particular, a \emph{city} is defined for each country to be a set of contiguous areas, each with a density of at least 1,000 people per square kilometer, yielding a total population of at least 10,000 (see Fig. \ref{fig:ua}).

For all countries except Japan, population count data was obtained for each 30''-by-30'' (approximately 1km-by-1km) grid from the LandScan (2015)$^{\text{TM}}$data base.%
\footnote{More specifically, we use the High Resolution Global Population Data Set copyrighted by UT-Battelle, LLC, operator of Oak Ridge National Laboratory under Contract No. DE-AC05-00OR22725 with the United States Department of Energy.}\ 
For Japan, population count data in 30''-by-45'' grids was obtained from the Grid Square Statistics of the 2015 Census of Japan.

The road distance between each pair of cities is computed as the shortest-path road distance between the most densely populated grids within each city
The road network data was downloaded from OpenStreetMap (\url{http://download.geofabrik.de/}).%
\footnote{Bilateral road distances are calculated by the Open Source Routing Machine (OSRM), which is an open-source routing engine designed for the geographic data of OpenStreetMap (\url{http://project-osrm.org/}).
More specifically, we used the routing service version 1 of OSRM with driving mode; the other settings of routing were taken from the default in OSRM.}\ 

For each country, we consider mostly its continental portion; however, if large islands are connected by roads or if reasonable road-equivalent distances can be computed, then these islands are included.
For example, Hainan in China and Hokkaido in Japan are included, while Hawaii in the US is not.

\section*{The Spacing-Out Property}

For our purposes, we first need to specify how a \emph{spatial} partition is constructed.
For any given set of cities for a given country, $\bm{U}$, and selection,$\{u_{1},\ldots,u_{K}\}$, of cities in $\bm{U}$, we first identify the subset, $\bm{U}_{i}$, of cities in $\bm{U}$ that are closest to each city, $u_{i}$, where ``closeness'' is here defined in terms of road distance between city locations.
This collection of subsets, $(\bm{U}_{1},\ldots,\bm{U}_{K})$ defines the \emph{Voronoi} $K$-\emph{partition} of $\bm{U}$ generated by these $K$ cities, where each subset, $\bm{U}_{i}$, is designated as a \emph{Voronoi} cell, and its size is defined by the number of cities in the cell.

For any given number, $L$, of the largest cities in $\bm{U}$, and for any partition, $v$, of $\bm{U}$, let $N_{L}(v)$ denote the number of partition cells of $v$ containing at least one of these $L$ cities.
If there is indeed substantial spacing between the largest cities in $\bm{U}$, then we would expect $N_{L}(v)$ to be larger for Voronoi partitions than for random partitions of the same cell sizes.
For given values of $L$ and $K$, we simulate $M\,\,(=1000)$ \emph{random} Voronoi $K$-partitions, $v=1,\ldots,M$, where the cities on which the Voronoi partitions are based are selected at random.
The resulting Voronoi count vector for these simulations is denoted by $\bm{N}_{L}\equiv [N_{L}(v):v=1,2,\ldots,M]$.%

For each of these Voronoi $K$-partitions, $v$, we then simulate $M$ random $K$-partitions, $\omega=1,\ldots,M$, of the same cell sizes. 
Note that these random partitions are formed without any regard to space.
Rather than conducting separate tests for each random Voronoi partition, $v$, which may produce rather uneven cell sizes, we construct a summary test statistic using appropriate mean values as follows.

We write the random partitions for $v$ as ordered pairs $(v,\omega)$, $\omega=1,\ldots,M,$ to indicate their size-dependency on $v$. 
In a manner paralleling $N_{L}(v)$, we then let $N_{L}(v,\omega)$ denote the number of cells in random partition $(v,\omega)$ that contain at least one of the $L$ largest cities in $\bm{U}$. In these terms the count vectors,  
\begin{equation}
\bm{N}_{L}(\omega)=[N_{L}(v,\omega)\,:\,v=1,\ldots,M]\,\,,\,\,\,\,\omega=1,\ldots,M\label{eq:random_counts}
\end{equation}
can each be regarded as random-partition versions of the Voronoi count vector $\bm{N}_{L}$.
In this setting, our basic null hypothesis is essentially that the Voronoi count vector, $\bm{N}_{L}$, is drawn from the same population as its random-partition versions in Eq.~\ref{eq:random_counts}.
But for operational simplicity, we focus only on the associated \emph{mean-counts}: 
\begin{equation*}
\overline{N}_{L}=\frac{1}{M}\sum_{v=1}^{M}N_{L}(v)
\end{equation*}
and 
\begin{equation*}
\overline{N}_{L}(\omega)=\frac{1}{M}\sum_{v=1}^{M}N_{L}(v,\omega)\,\,,\,\,\,\,\omega=1,\ldots,M\:.
\end{equation*}
In these terms, our explicit \emph{null hypothesis}, $H_{0}$, is that \emph{the Voronoi mean-count, $\overline{N}_{L}$, is drawn from the same population as its associated random mean-counts,} $\overline{N}_{L}(\omega),\,\,\omega=1,\ldots,M$.
If for the given set of simulated random partitions above, we now let $M_{0}$ denote the number of instances of $\overline{N}_{L}(\omega)$ which are at least as large as $\overline{N}_{L}$ (including the observed case itself), then the \emph{p-value}, $p_{0}$, for a one-sided test of $H_{0}$ is given by 
\begin{equation*}
	p_{0}=\frac{M_{0}}{M+1}\,.
\end{equation*}
\begin{figure}[h!]
\centering{}\includegraphics[scale=0.6]{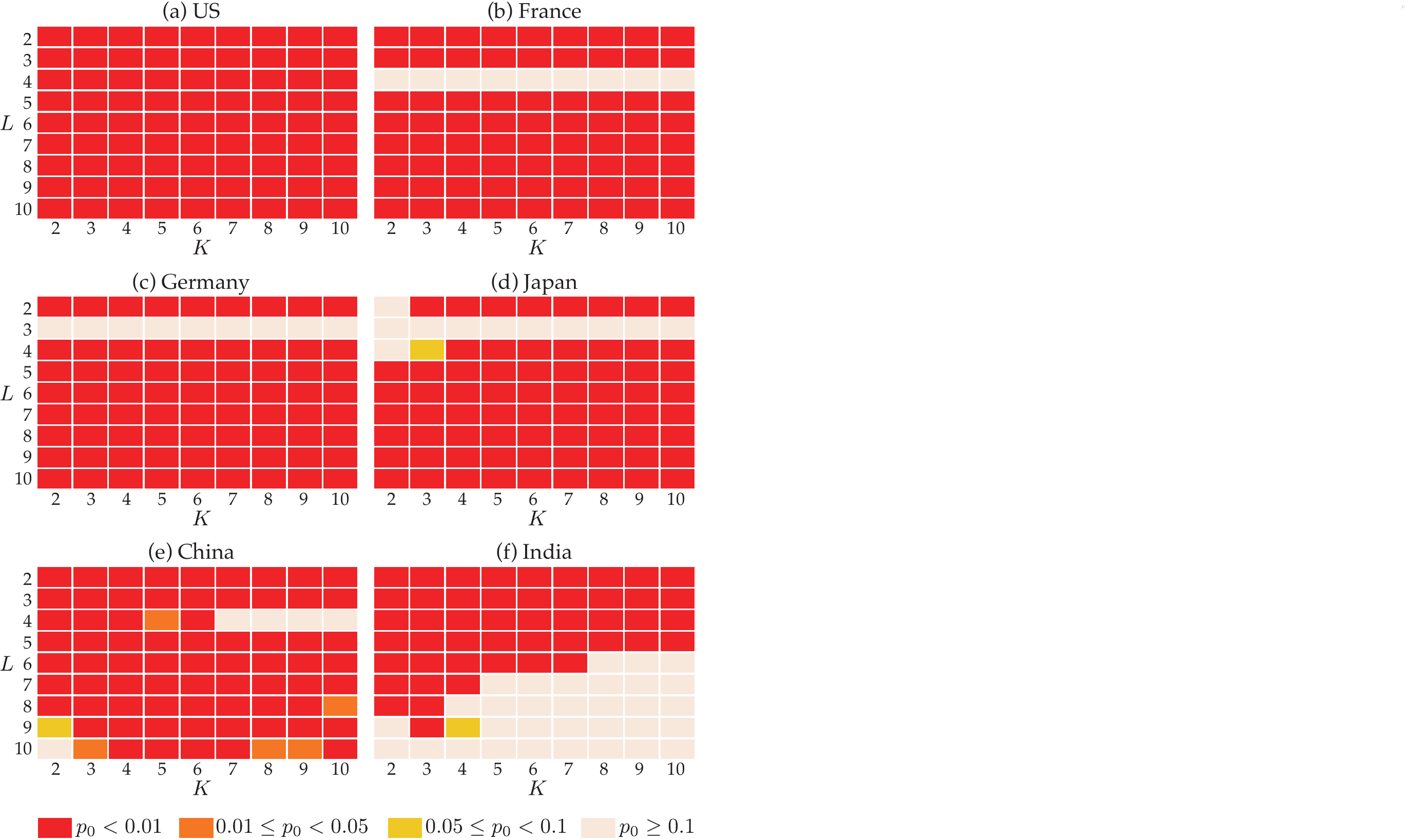}
\caption{Results of the spacing-out test\label{fig:spacing_10000}}
\end{figure}

The test results are given in Fig. \ref{fig:spacing_10000}. 
For each $K$ and $L$, the result is represented by red if $p_{0}<0.01$.
Similarly, orange, yellow, and linen colors indicate $0.01\leq p_{0}<0.05$, $0.05\leq p_{0}<0.1$, and $p_{0}\geq 0.1$, respectively.
Obviously, the evidence for US cities being spaced out is quite strong, as $p$-values are less than $0.01$ for all combinations of $K$ and $L$.
The evidence for France, Germany, Japan, and China is also quite strong except for a few cases.
For the case of France, the third and fourth largest cities (Marseille and Nice) are rather close; for Germany and Japan, the second and third largest cities (Essen and Cologne; Osaka and Nagoya) are rather close.
These indicate that natural geographic advantages matter for city locations, despite the fact that such advantages are to some degree controlled for by the construction of random Voronoi partitions;\footnote{Note that city sites are unevenly distributed in geographic space (think about plain versus mountainous areas).
In the construction of a random Voronoi partition, factors (such as natural advantages and economic development) that affect the density of city sites in a region are accounted for to some extent because the likelihood of each city site being drawn as a center of a Voronoi cell is the same. Hence, cities in regions with a high density of city sites are more likely to be drawn.} for example in the case of Japan, large flat areas are quite limited, and are mostly concentrated on the Pacific coast.
Nevertheless, the spacing-out property generally holds.
For India, the spacing-out property holds well up to and including the six largest cities, but not for cases where smaller cities are included.
Given India's current economic development, it is likely that locations of smaller cities are more influenced by natural geographic advantages.

\section*{Power Laws in City Size}

A city size distribution is said to satisfy a \emph{power law with exponent} $\alpha$ if and only if for some positive constant $c$ the probability of a city size $S$ larger than $s$ is given by
\begin{equation}
\Pr(S>s)\approx cs^{-\alpha},\quad s\rightarrow\infty.\label{eq:power-law}
\end{equation}
If a given set of $n$ cities is postulated to satisfy such a power law, i.e., with city sizes distributed as in Eq.~\ref{eq:power-law}, and if these city sizes are ranked as $s_{1}\geq s_{2}\geq\cdots\geq s_{n}$, so that the \emph{rank} $r_{i}$ of city $i$ is given by $r_{i}=i$, then it follows that a natural estimate of $\Pr(S>s_{i})$ is given by the ratio, $i/n\equiv r_{i}/n$. So by Eq.~\ref{eq:power-law} we obtain the following approximation, 
\begin{equation}
r_{i}/n\approx \Pr(S>s_{i})\approx cs_{i}^{-\alpha} \Rightarrow \ln s_{i}\approx b-\frac{1}{\alpha}\ln r_{i},\label{eq:zipf-plot}
\end{equation}
where $b=\ln(cn)/\alpha$. This motivates the standard log regression
procedure for estimating $\alpha$ in terms of the ``rank-size''
data, $[\ln(r_{i}),\ln(s_{i})]$, $i=1,\ldots,n$.

A natural way to estimate $\alpha$ is by running ordinary least squares on Eq.~\ref{eq:zipf-plot}.
However, many authors \cite{Gabaix-Ibragimov-2011,Nishiyama-et-al-2008}
have observed that this may underestimate $\alpha$ when smaller cities are included in the sample.
We use a simple procedure for correcting this bias, as proposed
by Gabaix and Ibragimov \cite{Gabaix-Ibragimov-2011}, by
subtracting 0.5 from the rank, which yields the modified regression, 
\begin{equation}
\ln(s_{i})=b-\theta\ln(r_{i}-0.5)+\varepsilon_{i}\,,\quad i=1,\ldots,n\label{eq:gi-regression}
\end{equation}
with $\theta=1/\alpha$.

\section*{Measuring the Commonality of Power Laws}

We now develop a method for examining the commonality of power laws for a collection of subsets of cities.
To do so, we start with an estimation of a model hypothesizing a common power law, and then develop an appropriate goodness-of-fit measure for this model.
Generally, if for any subset of cities, $\bm{U}_{j}\subseteq \bm{U}, j=1,\ldots,m$, it is true that the cities in each subset $\bm{U}_{j}$ exhibit power laws with a common exponent $\alpha$, then these subsets are said to exhibit the CPL.
Given the rank-size data for each subset $\bm{U}_{j}$, the regression framework in Eq.~\ref{eq:gi-regression} can be extended to a \emph{categorical regression} with fixed effects for each subset.
Let $n_{j}$ and $r_{ij}$ denote the number of cities and the rank of city $i$ in each subset $\bm{U}_j$.
Also let subset 1 denote a ``reference'' subset and for each other subset, $j=2,\ldots,m$, define indicator variables $\delta_{j}$ over the collection of subsets, $h=1,\ldots,m$, by $\delta_{j}(h)=1$ if $h=j$ and zero otherwise. 
For each $i$ and $j$, the desired categorical regression model is given by 
\begin{equation}
\ln s_{ij}=b_{1}-\theta\ln(r_{ij}-0.5)+\textstyle{\sum}_{h=2}^{m}\beta_{j}\delta_{j}(h)+\varepsilon_{ij}.\label{eq:cat-regression}
\end{equation}
Note that for any given subset $\bm{U}_j$ this model reduces to Eq.~\ref{eq:gi-regression}, where $b_{j}\equiv b_{1}+\beta_{j}$ for $j=2,\ldots,m$, and where the crucial slope coefficient $\theta$ (and hence $\alpha$) is the same for all subsets.

While the goodness of fit of this model can be measured in terms of \emph{R}-squared, one must then specify the joint distribution of the error terms, $\varepsilon_{ij}$, which in the present setting is completely unknown. 
However, our primary objective is not to gauge how well this model fits any given system, but rather to determine whether it yields a better fit for systems that are consistent with the spatial grouping property.
Hence our strategy is to use the least squares estimates of model in Eq.~\ref{eq:cat-regression} to construct a nonparametric goodness-of-fit measure, which is used to compare commonality of power laws between systems exhibiting spatial groupings and (appropriately defined) counterfactual systems that do not.

To do so, we start by using the least squares estimates $(\hat{\theta},\hat{b}_{1},\hat{\beta}_{2},\ldots,\hat{\beta}_{m})$ of the model parameters in Eq.~\ref{eq:cat-regression} to obtain the corresponding predictions, 
\begin{equation*}
\widehat{\ln s_{ij}}=\hat{b}_{1}-\hat{\theta}\ln(r_{ij}-0.5)+\sum_{h=2}^{m}\hat{\beta}_{j}\delta_{j}(h)
\end{equation*}
of log city sizes, $\ln s_{ij}$. 
While \emph{R}-squared could in principle still be used as a measure of fit in this nonparametric setting, there is general agreement that measures reflecting actual error magnitudes are more meaningful. By far the most commonly used measure of this type is \emph{root mean squared error} (RMSE). 
The RMSE for the estimated model above is given by: 
\begin{equation*}
\text{RMSE}=\sqrt{\frac{1}{\sum_{j=1}^{m}n_{j}}\sum_{j=1}^{m}\sum_{i=1}^{n_{j}}\big(\ln s_{ij}-\widehat{\ln s_{ij}}\big)^{2}}\,.\label{eq:rmse}
\end{equation*}
If the RMSE value for the given system is sufficiently small compared to those of the counterfactuals, then it can be concluded that this system is significantly more consistent with the CPL than are random counterfactuals.

\section*{Spatial Hierarchical Partitions}

Next, we develop specific collections of subsets of cities that are consistent with the spatial grouping property.
Note that when a Voronoi partition of the entire set of cities $\bm{U}$ is generated with the $L$ largest cities in $\bm{U}$ being the centers, then by construction, all cities are grouped around their closest large cities.
Thus, any such Voronoi $L$-partition of $\bm{U}$ is said to satisfy the \emph{spatial grouping property}.

If each cell of cities is taken to define a \emph{region}, then it is also reasonable to postulate that this relationship between large and small cities in each region is \emph{recursive}.
For example, suppose that San Francisco is included in the Los Angeles region.
Then in a similar manner, smaller cities around San Francisco might be included in a San Francisco subregion.
If so, then such relations generate a system of regions and subregions all exhibiting this same spatial grouping property.
Our interest is then in whether such systems also exhibit the CPL.

To be more specific, we now consider hierarchical regional systems consisting of many possible layers, where the subregions in each layer define Voronoi partitions of regions in the layer above. While there are a multitude of possibilities here, the simplest approach is to construct regional hierarchies with the same number of subregions in each region, as in central place theory dating from the seminal work of Christaller \cite{Christaller-1933}.
\begin{figure*}[t]
\centering{}\includegraphics[width=11.4cm, height=3.0656614cm]{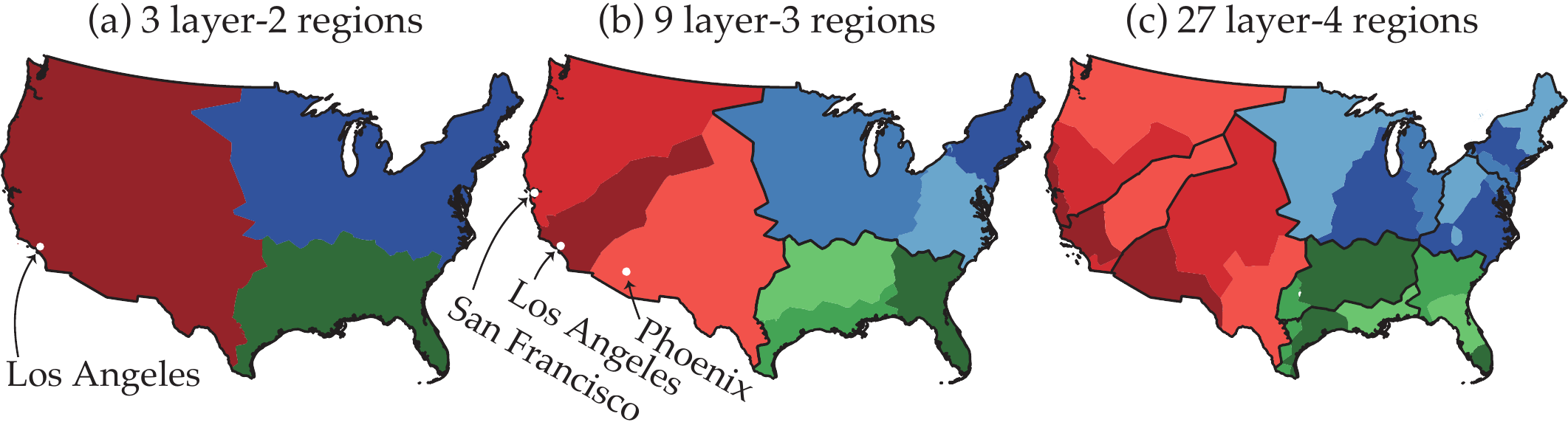}
\vspace{5pt}
 \caption{The spatial hierarchical 3-partition for the US. Here the partitions of land are based on Voronoi partitions of the set of cities with non-city land assigned to the closest cities.\label{fig:fractal_partition_us}}
\end{figure*}
\begin{figure*}[h!]
\centering{}\includegraphics[width=17.8cm,height=8.026807cm]{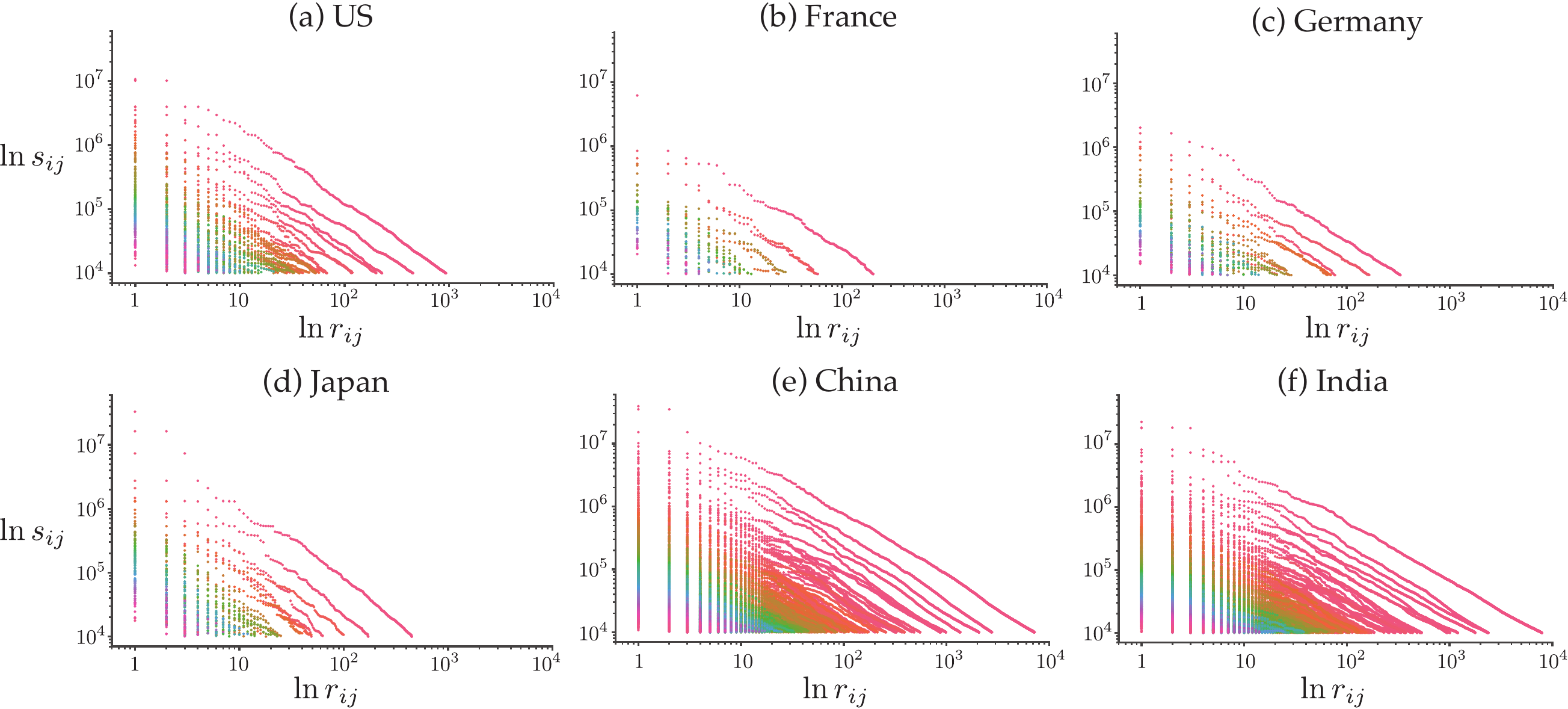}
\caption{For each country, the city size distribution in each cell of the spatial hierarchical three-partition is plotted. The numbers of cells constituting these hierarchical partitions are 368, 79,126, 181, 2,844 and 3,108 in the US, France, Germany, Japan, China and India, respectively.\label{fig:rs}}
\end{figure*}

This type of hierarchical system simplifies the analysis by allowing the number of subregions, $L$, to be left unspecified, so that tests can be conducted over a range of possible $L$ values.
Moreover, for each value of $L$, it allows a unique hierarchy of regions to be constructed that is fully consistent with the spatial grouping property.

The construction of these hierarchies is simple. To compare possible power laws for the country as a whole with those of its subregions, we start by treating the country itself as a region \textendash{} which by definition exhibits spatial grouping with respect to its largest city.
For any given $L$, we then choose the $L$ largest cities in the country (including the largest city) and take these to define the \emph{central cities} for a Voronoi $L$-partition of the country region.
This yields a two-layer hierarchy consisting of the country region and $L$ subregions. 
This hierarchy is then extended by choosing the $L$ largest cities in each subregion (including its central city) and defining a new Voronoi $L$-partition of subregions with respect to these central cities.
Of course this process cannot be continued indefinitely, since there are only finitely many cities in a country. So our ``stopping rule'' is that no region can be divided into $L$ subregions if it contains less than $L$ cities.
This process results in a unique hierarchical partition which reflects the spatial grouping property at every layer, and is thus designated as a \emph{spatial hierarchical $L$-partition.}

As an example, we now consider the spatial hierarchical 3-partition for the US.
The first layer of this system, associated with the largest city (New York), is by definition the whole country.
The second layer constitutes the Voronoi 3-partition generated by the three largest cities in the US (New York, Los Angeles, Miami), as shown in Panel (a) of Fig. \ref{fig:fractal_partition_us}.
The third layer shown in Panel (b) then consists of three Voronoi 3-partitions, each generated by the three largest cities in one of the Voronoi regions in the second layer.
For example the three largest cities in the New York region (New York, Chicago, and Washington D.C.) define the relevant third-layer partition of this particular region.
Panel (c) further shows 27 layer-4 regions.

Note for example that New York is by definition the central city of one region in each layer.
If these regions are viewed as successively more local hinterlands of New York, then it is natural to designate the largest of these (i.e., the highest-layer Voronoi region in which New York appears as the central city) as the \emph{global hinterland} of New York.
For New York in particular, this global hinterland is the entire country.
Similarly, the global hinterland of the second-layer city, Los Angeles, is shown by the red region in Fig. \ref{fig:fractal_partition_us}(a), and that for the third-layer city, Phoenix, is shown by the light red region in Fig. \ref{fig:fractal_partition_us}(b). 
Since the size of each of these cities is more directly related to its global hinterland than to any of its local hinterlands, we now designate the city size distribution of its global hinterland as the \emph{city size distribution for that city}.
With these conventions, the city size distributions for every central city in the spatial hierarchical partition for each country are shown in Fig. \ref{fig:rs}.

Here power laws appear to be good approximations of the rank-size data, and there seems to be reasonable agreement between the slopes of these curves. 
But to test the significance of the commonality of power laws, and in particular, to isolate the contribution of spatial grouping, it is necessary to construct a statistical population of random hierarchical partitions that differ from this given system only in terms of spatial grouping.

To do so, we replace ``largest-city Voronoi $L$-partitions'' with ``largest-city random $L$-partitions'' in the sense that cities are assigned to the $L$ largest cities randomly and hence without any regard to spatial relations.
This process of generating largest-city random $L$-partitions is repeated recursively in each cell with the constraint that the sizes of cells in each layer are given by the actual spatial hierarchical partition.

\section*{Testing the Spatial CPL}

To perform the actual tests for any value of $L$, we begin by generating $N=1000$ random hierarchical $L$-partitions.
For any given $L$, the categorical regression in Eq.~\ref{eq:cat-regression} can be estimated for both the observed spatial and random hierarchical $L$-partitions. 
This estimation procedure will then yield an RMSE$_{L}$ value for the observed spatial hierarchical $L$-partition together with RMSE$_{Lv}$ value for each of the simulated random hierarchical $L$-partitions, $v=1,\ldots,N$.
In this context, the relevant null hypothesis to be tested is that the observed spatial hierarchical partition is simply another instance of these random hierarchical partitions. 
Thus the effective sample size under the null hypothesis is $N+1$.
So if we now let $N_{L}$ denote the number of RMSE$_{Lv}$ values not exceeding RMSE$_{L}$ (including the observed case itself), then the fraction 
\begin{equation*}
p_{L}=\frac{N_{L}}{N+1}\label{eq:pval for R_square-1}
\end{equation*}
is the estimated \emph{p}-value for a one-sided test of this null hypothesis.
\begin{figure}[h!]
\centering{}\includegraphics[scale=0.6]{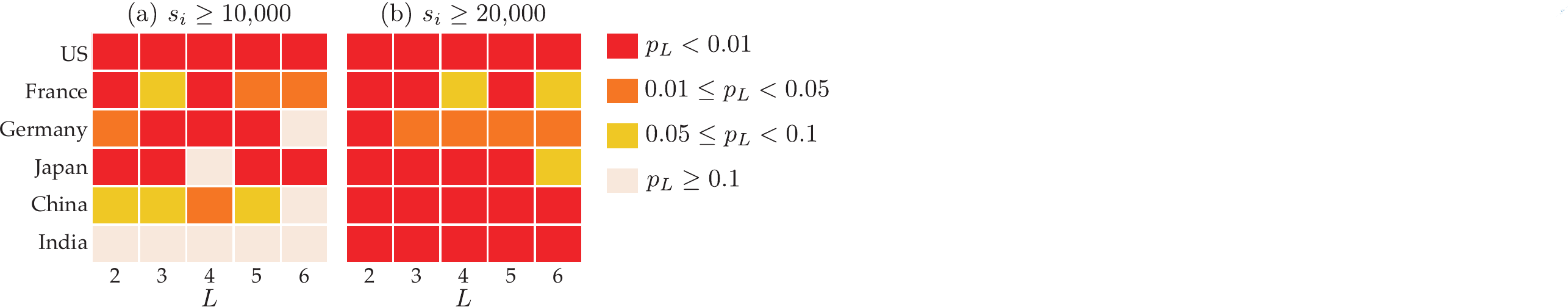}
\caption{Results of the Spatial CPL Test. Panels (a) and (b) show the results when a city is defined to have at least a population of 10,000 and 20,000, respectively. \label{fig:cpl}}
\end{figure}

The test results are shown in Fig. \ref{fig:cpl}.
First, it is clear from Panel (a) that the CPL under spatial grouping, i.e., the \emph{spatial CPL},  holds very tightly for the US, since the likelihood of random counterfactuals exhibiting stronger CPL properties than the observed spatial hierarchical $L$-partition for the US is less than $0.01$ for all values of $L=2,\ldots,6$.
For France, Germany, and Japan, the spatial CPL also holds quite significantly overall.
For the case of China, it is significant but to a lesser degree.
Only for the case of India does the spatial CPL fail to be significant for any value of $L$.

We conjecture that this lack of significance is related to low degree of urbanization in India, given its current stage of economic development. Moreover, the high level of overall population density in India suggests that even in rural areas the local density of population may often be sufficiently high to qualify as ``cities'' under our definition above. In particular, it can be seen by a close examination of the India map in
Fig. \ref{fig:ua} that the Ganges Basin is filled with ``cities''.
Similar observations can be made for China (perhaps with a lesser degree).
Compared with the number of cities in the US (931), the numbers of cities in India and China are much larger at 7,913 and 7,203, respectively, while the populations of these two countries are only roughly four times as large as the US.

To check these observations further, we next modified our definition of cities by increasing the total population threshold to be at least 20,000 inhabitants. Under this more stringent definition, the numbers of cities in both  India and China are essentially cut in half (3,480 and 3,524, respectively). More importantly, a repetition of the above analysis under this city definition shows [Fig. \ref{fig:cpl}(b)] that the spatial CPL for India and China now holds very tightly, as well as for the US.
For France, Germany, and Japan, the smaller total areas of these countries together with this more stringent city definition effectively reduces the number of cities (samples) to the point where categorical regression results are affected. Nonetheless, the spatial CPL for these countries continues to be relatively significant. Note particular that Panel (b) now exhibits no insignificant cases.  
\begin{figure}[h!]
\centering{}\includegraphics[scale=0.6]{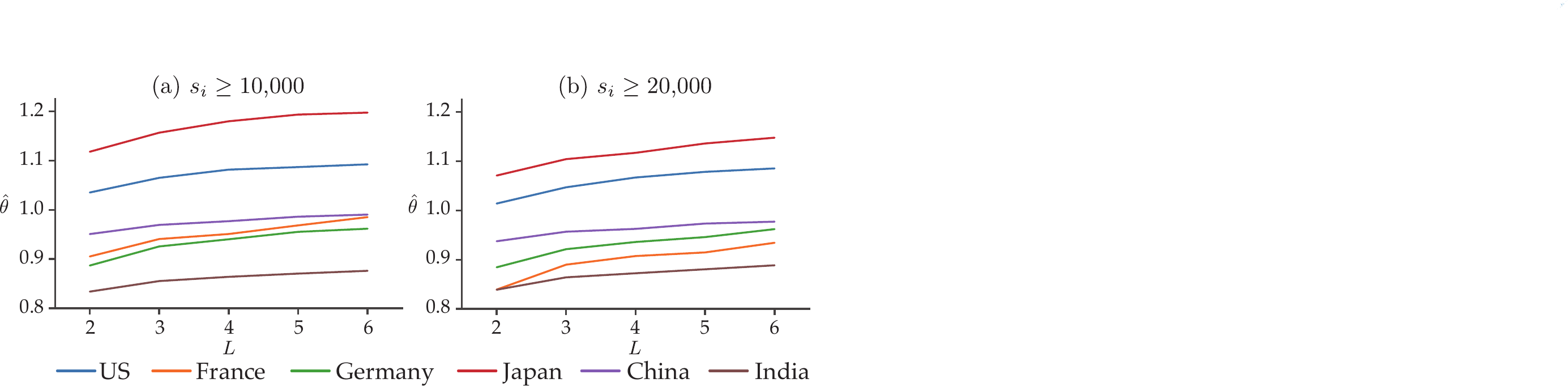}
\caption{The estimated common slope coefficient $\hat{\theta}$ in Eq.~\ref{eq:cat-regression}. Panels (a) and (b) show the results when a city is defined to have at least a population of 10,000 and 20,000, respectively.\label{fig:cpl_slope}}
\end{figure}

Taken together, the results in Fig. \ref{fig:cpl} provide strong evidence for the spatial CPL in all countries we have tested. These findings in turn raise the question of whether there might be a ``common'' power law for all these countries.
Fig. \ref{fig:cpl_slope} shows the estimated CPL exponents $\hat{\theta}$ for each country over a range of $L$ values. 
Note in particular that the between-country variations in $\hat{\theta}$ values are substantially larger than their within-country variations across $L$. This suggests that a common power law across countries is unlikely.

\section*{Conclusion}

Using data from China, France, Germany, India, Japan, and the US, we first document the spacing-out property that large cities are much more spaced out than their random counterparts. Given the ubiquity of smaller cities and towns, this suggests the existence of local city systems surrounding the largest cities and thus supports the spatial grouping property. Using the same data, spatial hierarchical partitions are formed, and it is found that city size distributions in different parts of this hierarchical structure exhibit a high degree of commonality in terms of power laws compared with their random counterparts. This spatial CPL suggests the existence of a spatial fractal structure.

An alternative explanation of the CPL for countries is suggested by the theory of random growth processes, as in \cite{Gabaix-1999} and related literature. However, this theory implies that the CPL should hold for essentially all random subsets of cities within a country, and thus should hold for our random counterfactuals. But our test results suggest that the CPL is much stronger for spatial hierarchical partitions of cities than for random subsets, and thus cast doubt on this random growth explanation. 

More generally, our results point toward theories that generate city systems as spatial fractal structures.
One prominent candidate is central place theory \cite{Christaller-1933,Beckmann-1958,Hsu-2012,Fujita-Krugman-Mori-1999,Tabuchi-Thisse-2011}.
The central tenets of this theory assert that the heterogeneity of goods together with the spatial extent of markets gives rise to natural hierarchies of cities, and thus to a diversity of city sizes.
The resulting central place hierarchies, as depicted by \cite{Christaller-1933,Hsu-2012}, are clearly spatial fractal structures.
In fact, the model in \cite{Hsu-2012} generates a spatial CPL, although relying on more complex structural assumptions than standard fractal theories \emph{\`{a} la} \cite{Mandelbrot-1982}.
An alternative approach, somewhat closer to standard fractal theories, utilizes insights from central place theory to develop city systems as fractal structures based in city-hinterland relations \cite{Batty-Longley-1994}. But such systems are not yet sufficiently explicit to draw conclusions about spatial CPL properties.

Another possibility is to extend random growth processes by adding spatial relations among cities \cite{Eaton-Eckstein-1997,Rosenfeld-et-al-2008,Rybski-Ros-PHYR-E2013}, and to examine the performance of these processes in terms of spatial CPL.
A final possibility is to adopt techniques from spatial networks \cite{Clauset-Moore-Newman-2008,Gastner-Newman-2006} to develop a theory of city systems. 
While such network concepts have been applied to link different locations within a given city \cite{Batty-Longley-1994,Batty-2013}, it remains to be determined how these concepts might be generalized to a city system. 




\showmatmethods{} 

\acknow{This research was conducted as part of the project, ``Agglomeration-based Framework for Empirical and Policy Analyses of Regional Economies'', undertaken at the Research Institute of Economy, Trade and Industry. 
This research has also been supported by the International Joint Research Center of Advanced Economic Theory of the Institute of Economic Research in Japan, and  the Grant in Aid for Research (Nos. 17H00987, 16K13360, 16H03613, 15H03344) of the MEXT, Japan. A substantial portion of the research work was conducted when the authors visited the Institute of Economics of Academia Sinica, and their hospitality is appreciated. We also thank Yu-Chih Luo for his excellent research assistance.}

\showacknow{} 

\bibliography{CPL.bib}

\end{document}